# Radiation from Kinetic Poynting Flux Acceleration

*Edison Liang[1] and Koichi Noguchi[1]*

## ABSTRACT

We derive analytic formulas for the power output and critical frequency of radiation by electrons accelerated by relativistic kinetic Poynting flux, and validate these results with Particle-In-Cell plasma simulations. We find that the in-situ radiation power output and critical frequency are much below those predicted by the classical synchrotron formulae. We discuss potential astrophysical applications of these results.

*Subject Heading*s: Acceleration of particles– Radiation mechanisms: non-thermal - Gamma-rays:bursts

*Online Material*: color figures

## 1. INTRODUCTION

In popular paradigms of radiation from blazars, pulsar wind nebulae (PWN), gamma-ray bursters (GRB) and other gamma-ray sources, relativistic outflow energy (hydrodynamic or electromagnetic) from the central compact object (black hole or neutron star) is first converted into relativistic nonthermal electrons via some collisionless dissipation mechanisms (e.g. shocks, Dermer 2003, Meszaros 2002, Lyubarski 2005). These nonthermal electrons are then hypothesized to radiate synchrotron-like radiation, including small-pitch-angle synchrotron (Epstein and Petrosian 1973, Lloyd and Petrosian 2000), or "jitter" radiation if the magnetic field is too chaotic (e.g. due to Weibel instability, Weibel 1958, Medvedev 2000, Medvedev et al 2005). In addition, inverse Comptonization of the synchrotron photons (SSC) or external soft photons (EC) may account for the high-energy (e.g. MeV-TeV) gamma-rays (Dermer et al 2000, 2003). Most popular astrophysical models invoke the classical synchrotron formulas (Rybicki and Lightman 1979). However, two outstanding questions remain unsolved: (a) exactly how is the outflow energy converted into nonthermal electron energy via collisionless shocks (CS, Hoshino et al 1992, Gallant et al 1992, Silva et al 2003, Nishikawa et al 2003, Spitkovski 2006), or electromagnetic Poynting flux (PF, Smolsky & Usov 2000, Lyutikov & Blackman 2002, Van Putten & Levinson 2003, Lyutikov and Blanford 2003)? (b) do the accelerated electrons always radiate synchrotron radiation, since the synchrotron models do not agree with observations in many cases (Fenimore 2002, Dermer & Chang 1999, Preece et al 2000)? In this paper we present concrete examples of acceleration mechanisms whose radiation process is drastically different from classical synchrotron radiation.

Over the past few years we have used sophisticated Particle-in-Cell (PIC) codes for relativistic collisionless plasmas (Langdon and Lasinksi 1976, Birdsall & Langdon 1991, Langdon 1992) to study nonthermal electron acceleration and radiation processes (Liang et al 2003, Liang & Nishimura 2004, Nishimura et al 2003, Liang & Noguchi 2005, 2006). A unique feature of our PIC simulations is that the power radiated in-situ by each superparticle (=numerical representation of a charged particle) can be computed simultaneously as the superparticles are accelerated by the Lorentz force (Noguchi et al 2005, Liang and Noguchi 2005, 2006). This approach provides a fully *self-consistent treatment of the intrinsic radiation output during the acceleration process*. In this paper we focus on the radiation of plasmas accelerated directly by intense electromagnetic pulses or Poynting flux (PF), and derive analytic formulas for this radiation from first-principles. Section 2 reviews the basic physics of PF acceleration. Section 3 briefly summarizes the key result of the numerical radiation power. In Section 4 we derive the critical frequency of PF radiation. In Section 5 we derive an analytic formula for the radiation power output. In Section 6 we speculate on the astrophysics scenarios of PF acceleration. In Section 7 we apply the analytic formulas to a sample PF model of long GRBs. Section 8 is devoted to discussion and summary.

A common misconception about PIC simulations is that such simulations are too small in physical scale (measured in units of plasma skin depths and electron gyroradii) to be relevant to macroscopic astrophysical phenomena. However, unlike MHD simulations, the purpose of PIC simulations is not to try to reproduce macroscopic phenomena, but to *discover and quantify microphysical laws* governing particle energization, radiation mechanisms, wave-particle interaction and dissipation processes, which operate at the level of plasma skin depths and gyroradii. Once discovered via numerical simulations, such physical laws should be rederived analytically from first principles. These validated laws are then applicable to macroscopic phenomena irrespective of the space and time scales. This is the approach we will adopt in this paper.

## 2. ACCELERATION BY KINETIC POYNTING FLUX

In this paper we are interested in relativistic collisionless plasmas whose Coulomb mean free paths are much larger than the relevant plasma scale sizes (see Sec.7 for sample numbers). The interaction of such plasmas with intense EM fields lies outside the classical MHD regime and can only be treated correctly with kinetic theory. Here we define "Poynting flux" as a directed plasma outflow dominated and driven kinetically by *transverse* electromagnetic (EM) fields with $\Omega_e/\omega_{pe}=B/(4\pi nm)^{1/2} >1$, ($\Omega_e = eB/m$ =electron gyrofrequency, $\omega_{pe} =(4\pi ne^2/m)^{1/2} =$ electron plasma frequency, m=electron mass, n=imbedded electron density, c = 1 throughout this paper except in Sec.7), in the absence of a flow-aligned longitudinal magnetic field. Hence we will not consider diffusive particle acceleration by scattering with classical Alfven and whistler waves (see discussions in Sec.8) in a background magnetic field, or electrostatic acceleration by longitudinal plasma (Langmuir) waves (Boyd and Sanderson 1969).  Instead we focus on particle acceleration by the ponderomotive (**JxB**) force.  Astrophysical examples of such kinetic electromagnetic outflows include the equatorial stripe wind of pulsars and magnetars (Lyubarsky 2005, Skjaeraasen et al 2005), and the front end of a low-density magnetic tower jet driven by magnetized accretion disks around black holes (Koide et al 2004).

There are two opposite situations in which a kinetic PF can efficiently transfer its EM energy into the nonthermal kinetic energy of a plasma: pushing or pulling the plasma with the **JxB** force (Fig.1).  The common "pushing" mechanism involves an intense EM pulse impinging on an overdense plasma ($\omega_{pe} > 2\pi/\lambda$, $\lambda$=characteristic wavelength of the EM field) from the outside (Fig.1a).  This phenomenon is well known from the interaction of intense radio waves with the ionosphere, or from laser-plasma interactions (Kruer et al 1975).  The intense EM field induces a ponderomotive (**jxB**) force near the critical surface (where $\omega_{pe} = 2\pi/\lambda$, Fig.1a), which accelerates the surface electrons inward to Lorentz factors characterized by the local dimensionless vector potential $a_o$ ($= eB\lambda/2\pi m_e$, Wilks et al 1992).  When $a_o \gg 1$, the EM field "snowplows" all upstream electrons.  The reflection front moves forward relativistically so that the EM pulse suffers little reflection (Kruer et al 1975).  In extreme cases (e.g. in an e+e- plasma) the bulk of the EM energy is transferred to relativistic particles instead of being reflected.  However, this type of "pushing" PF acceleration mainly produces quasi-Maxwellian "superthermal" electron spectra with $kT \sim a_o$ (Fig.2a, Wilks et al 1992), instead of the power-law spectra (Preece et al 2000) commonly observed in astrophysics.  We call this "pushing" acceleration by an intense EM pulse "leading Poynting or ponderomotive acceleration" or LPA (Fig.1a).  PIC simulations and analytic theory (Harteman and Kerman 1996) suggest that the maximum Lorentz factor achievable by LPA is limited to $\max(\Omega_e^2/\omega_{pe}^2, a_o^2/2)$ due to energy and momentum conservation, since all upstream electrons must share the PF momentum and energy. However, because the snowplow is moving at almost light speed and the electrons are highly collisionless, in most cases no forward shock (in the conventional sense) is observed to form in the upstream plasma. Astrophysically, LPA is potentially relevant to the interaction of pulsar or magnetar winds with a dense environment in the kinetic limit.

In contrast to the LPA, the "trailing Poynting or ponderomotive acceleration" or TPA, occurs when an intense EM pulse pulls, instead of pushes, an overdense plasma (Fig.1b, TPA replaces the acronym DRPA used in our early publications, Liang et al 2003, Liang & Nishimura 2004). Consider for example a situation in which a strongly magnetized, overdense collisionless plasma with $B/(4\pi nm)^{1/2} > 1$ suddenly expands due to force imbalance. The expansion disrupts the sustaining current, so that $4\pi \mathbf{J} < \mathbf{Curl\ B}$. In the absence of an external EMF regenerating the current, the excess displacement current ($\partial \mathbf{E}/\partial t$) then generates a transverse EM pulse, which tries to escape from the embedding plasma (Fig.1b). As the EM pulse emerges from the plasma, it "pulls" out the surface electrons via the $\mathbf{jxB}$ force, where $\mathbf{j}$ is the self-induced polarization current (Boyd and Sanderson 1969). When the $\mathbf{jxB}$ force is ultra-intense, the accelerated electrons can stay *comoving* with the *group* velocity of the EM pulse which is < c due to plasma loading, and the acceleration can be sustained. Unlike the LPA case in which the maximum Lorentz factor is limited by momentum conservation, TPA transfers the PF energy and momentum to a *decreasing* number of fast electrons, as slower electrons continuously fall behind the pulse (Liang & Nishimura 2004). We find that the maximum Lorentz factor achieved by TPA is unlimited until radiation damping or dephasing (e.g. due to wave-front curvature) become important. PIC simulations show that TPA always accelerates the high-energy electrons into a simple power law independent of the initial conditions or the pulse width (Fig.2b). The TPA mechanism is exceedingly robust and efficient, typically converting > 50% of the EM energy into accelerated particle energy over a short distance(see Sec.7).

Physically both LPA and TPA are caused by a relativistic **E x B** drift in which the transverse EM field comoves with the particle drift velocity. The key difference is that in LPA the plasma load snowplowed by the EM pulse increases or stays constant with time, thereby limiting the Lorentz factor, whereas the TPA Lorentz factor increases indefinitely due to decreasing plasma loading (Liang et al 2003, Liang and Nishimura 2004). Density-wise, LPA involves collisionless compression of the plasma, while TPA involves rarefaction of the plasma. In contrast to shocks, in which bulk flow energy is converted into EM energy, either via compression of upstream fields or via Weibel (1959) and other instabilities, LPA and TPA converts EM energy into accelerated particle energy. We emphasize that both LPA and TPA are strictly kinetic phenomena with no analog in the MHD limit. The detailed physics of LPA and TPA has been reviewed extensively elsewhere (Liang and Noguchi 2005, 2006), so they will not be repeated here.

TPA may be relevant to radiation in astrophysics at two different levels: global and local. Globally, EM pulses with large-scale ordered fields may be generated due to reconnection of a magnetic tower jet or magnetar wind (Koide et al 2004), or from the merger of strongly magnetized neutron stars into a black hole. For example, TPA can take place when the front end of a disconnected magnetic tower jet propagates down the steep density gradient of the collapsar envelope and turns into an unconfined kinetic EM pulse (see Sec.7). Similarly, when a millisecond magnetar or a merging strongly magnetized neutron star binary collapses to form a black hole, part of its energy may be emitted in the form of an intense EM pulse.

However, TPA may also occur at the local level even in the absence of large scale ordered EM fields. For example, relativistic EM turbulence generated by shocks and shear layers may dissipate via the TPA mechanism as the nonlinear waves propagate into low density regions with $\Omega_e/\omega_{pe} > 1$. In this case sustained comoving particle acceleration can last locally until dephasing occurs. We emphasize that the only piece of physics invoked in the radiation calculation below is that the particle is *accelerated locally by a nearly comoving transverse EM wave*. No assumption about the global geometry, topology or size scale of the EM field is required. Hence the radiation formulas derived in this paper have more general validity and much broader applications than the small scale PIC simulations performed so far based on the simplistic LPA and TPA scenarios (Liang and Noguchi 2005, 2006).

## 3. NUMERICAL RADIATION POWER OUTPUT

In this paper we focus on the radiation output of electrons (and positrons) accelerated by a comoving kinetic Poynting flux. Numerically, we compute the radiation power output by incorporating the relativistic dipole formula (Rybicki & Lightman 1979) into our PIC code:

$$P_{rad} = 2e^2 \, (F_{\parallel}^2 + \gamma^2 F_{+}^2)/3m^2 \tag{1}$$

where $\gamma$ = Lorentz factor, $F_{\parallel}$ = force component along velocity $\mathbf{v}$, and $F_{+}$ = force component orthogonal to $\mathbf{v}$. We compute the total power loss of each superparticle in the PIC simulation, by interpolating the field data from the cell boundaries to the instantaneous superparticle position, so that $\mathbf{F}$ and $\mathbf{v}$ refer to the *same time and space* point. We have carefully calibrated this numerical procedure with known analytic results. Fig.3 compares the numerical radiation output for an isotropic thermal plasma in a static uniform B field with that computed using the analytic synchrotron formula (Rybicki and Lightman 1979). Their excellent agreement, especially for the high-energy electrons, validates our numerical algorithm. However, PIC simulation cannot be used to compute the radiation spectrum numerically because the PIC simulation time step (typically = 0.25 gyroperiod) is too large to accommodate the high frequencies.

The upper panels of Figs.4 & 5 illustrate the evolution of the $P_{rad}$ distribution of superparticles in sample LPA and TPA runs.  In both cases a plane, linearly polarized EM pulse accelerates the same slab of overdense e+e- plasma, one from the outside and one from the inside.  While the energies of the pairs increase monotonically due to the acceleration, the power radiated by the electrons initially rises to a maximum, but then declines monotonically.  We find that in both cases $P_{rad}$ << the classical synchrotron power $P_{syn} = 2e^4B^2p_{+}^2/3m^2$ (m$p_{+}$=momentum orthogonal to **B**).  This suppression of radiative power can be understood as follows.  As particles are accelerated to higher and higher $\gamma$, **v** aligns increasingly with the Lorentz force **F**.  So the $F_{+}$ term in Eq.(1) decreases relative to the $F_{\parallel}$ term.  However, $P_{syn}$ comes only from the $F_{+}$ term (Rybicki & Lightman 1979).  Hence $P_{syn}$ >> $P_{rad}$ for high $\gamma$ particles.  In Sec.5 we will rigorously derive a general analytic formula for $P_{rad}$.  Here we will only state that, for electrons almost comoving with the EM pulse, $P_{rad}$ can be approximated by:

$$P_{analytic} = P_{syn} \sin^4\alpha \qquad\qquad\qquad (2)$$

where $\alpha$ is the angle between $p_{+}$ and the Poynting vector **k** (Fig.6).  Fig.7 compares the numerical $P_{rad}$ with $P_{analytic}$ for the runs of Figures 4 & 5.  It shows good correlation for the high-$\gamma$ particles. Since $\sin\alpha$ <<1 for high-$\gamma$ particles (c.f. Fig.8), Eq.(2) explains why $P_{rad}$ << $P_{syn}$.  The rise and decline of $P_{rad}$ in Figs.4 & 5 are caused by the competition between increasing $\gamma$ (and $P_{syn}$) and decreasing $\alpha$.

## 4. RADIATION CRITICAL FREQUENCY

A prominent feature of GRB and blazar spectra is the presence of a low energy spectral break $E_{pk}$ (hundreds of keV for classical GRBs, radio-IR for blazars).  This spectral break is an indicator of the overall spectral hardness, and is usually interpreted as the critical frequency of synchrotron radiation $\omega_{crsyn} \sim 1.5\Omega_e\gamma_o p_{+o}$ (Rybicki & Lightman 1979) by electrons with low energy cutoff $\gamma_o$. This interpretation of the spectral break, together with some assumptions about energy equipartition, is often used to constrain the Lorentz factor and magnetic field of the source. However, as we show below, for radiation emitted by TPA and LPA electrons, the asymptotic critical frequency $\omega_{cr}$ is $<< \omega_{crsyn}$.

To derive the formula for $\omega_{cr}$, we follow the approach of Landau and Lifshitz (1980): $\omega_{cr}$ is determined by the time (measured in detector frame) it takes the radiation beam of opening angle $1/\gamma$ to sweep past the detector due to the curvature of the particle trajectory. For electrons comoving or almost comoving with the PF, the parallel momentum $p_x$ (x is the direction along **k**, Fig.6) increases monotonically while $p_z$ (momentum along **E**) asymptotes to a constant (Liang & Nishimura 2004, note that $p_y$ along **B** is constant to first order). Hence the change in the radiation beam direction due to bending of particle trajectory is dominated by the change in $p_x$: $\Delta\theta \sim p_z\Delta p_x/p_x^2$. From the Lorentz force equation we have $d\gamma/dt=eE_zp_z/m\gamma$. Hence the time in the laboratory frame for the radiation beam to change by an angle $\Delta\theta \sim 2/\gamma$ is $\Delta t=2\gamma^2 m/(eE_zp_z^2)$ where we have used the approximation $\gamma \sim p_x (>> p_z, p_y)$. This translates into a duration in the detector frame $\Delta t_{ob}=\Delta t/2\gamma^2=m/eE_zp_z^2$. Thus the critical frequency (Rybicki and Lightman 1979):

$$\omega_{cr}=1.5/\Delta t_{ob}=1.5eE_zp_z^2/m=1.5\Omega_e p_z^2=1.5\Omega_e \ p_+^2\sin^2\alpha \sim \omega_{crsyn}\sin^2\alpha \qquad (3).$$

Since $\sin\alpha << 1$ at high $\gamma$ (Fig.8), $\omega_{cr} << \omega_{crsyn}$  In Sec.6 we will discuss the major implications of this result for modeling GRB and blazar data. .  In the lower panels of Figs.4 & 5 we plot the evolution of $\omega_{cr}$ of the same LPA and TPA runs.  These snapshots highlight the evolution of the spectral hardness.  Again we see that $\omega_{cr}$ first rises to a maximum before declining monotonically due to the competition between increasing $\gamma$ and decreasing $\alpha$. However the decline of $P_{rad}$ is more rapid than $\omega_{cr}$ due to the extra factors of $\sin\alpha$.

## 5. RADIATION POWER FORMULA FOR KINETIC PF ACCELERATION

In this section we derive a general analytic approximation for the power radiated by an electron accelerated locally by a comoving kinetic PF. While the following derivation assumes linearly polarized plane waves for simplicity, the results should be valid in general 3D geometry as long as the wave front curvature and transverse gradients are << 1/(acceleration distance) (distance for e-folding increase in $\gamma$). We emphasize that this radiation formula should be applicable to any acceleration by transverse EM fields almost comoving with the local **ExB** drift velocity. Hence its potential applications in astrophysics are much broader than the specific LPA or TPA scenarios discussed above.

For particle motion in a linearly polarized plane wave with $(\mathbf{E},\mathbf{B}) = (E_z, B_y)$ (Fig.6), we have $F_x=-ev_zB_y$; $F_y=0$; $F_z=e(E_z+v_xB_y)$. Here x is the direction of Poynting vector $\mathbf{k}$. After a little algebra we find:

$$F_{||}=eE_zv_z/v; \quad F_{\perp}^2=e^2B_y^2[\sin^2\alpha(v^2-v_w^2)+(v_x-v_w)^2] \tag{4}$$

where $v_w=-E_z/B_y$ is the local profile speed of the EM field ($v_w<1$ due to plasma loading) and $\sin\alpha=v_z/v$. Substituting Eq.(4) into Eq.(1) we obtain:

$$P_{analytic} = 2e^4B_y^2[\sin^2\alpha(\gamma^2-1)(1-v_w^2)+\gamma^2(v_x-v_w)^2]/3m^2 \tag{5}.$$

Hence the power radiated by a PF-accelerated electron depends in general on two key parameters: the local EM field profile speed $v_w$ and the angle $\alpha$ between velocity $\mathbf{v}$ and Poynting vector $\mathbf{k}$. Eq.(5) simplifies in various special limits:

A. Comoving particles ($v_x=v_w$):

In this case Eq.(5) simplifies to $P_{analytic} = 2e^4B_y^2(p_z^2+p_y^2)sin^2\alpha/3m^2$ when $\gamma >> 1$. In all of our runs, $p_z >> p_y$ at late times. So this reduces to

$$P_{analytic} = 2e^4B_y^2p_+^2sin^4\alpha/3m^2$$

which is Eq.(2). As Fig.7 shows, Eq.(2) is a good approximation for electrons comoving with $v_w$. However, Eq.(2) is not a good approximation for electrons out of phase with $v_w$ (note that electrons can have $v_x > v_w$ or $v_x < v_w$). Liang and Nishimura (2004) suggested that $v_w$ corresponds roughly to the peak $\gamma$ of the particle distribution function $f(\gamma)$.

B. Vacuum pulse limit ($v_w=1$):

In the limit $v_w=1$, the PF propagates as a vacuum EM pulse, Eq.(5) becomes for $\gamma >> 1$:

$$P_{analytic} = 2e^4B_y^2\gamma^2 (1-v_x)^2/3m^2 \sim e^4B_y^2\gamma^2sin^4\alpha/6m^2 \qquad (6)$$

Since $p_+ \sim \gamma$, Eq.(6) has the same form as Eq.(2) but is a factor of 4 less in magnitude. It defines the *lower limit* to the radiative power loss of a PF accelerated electron since in reality $v_w<1$.

C. Slightly subluminal PF ($1-v_w=\varepsilon<<1$)

For most astrophysics applications, the PF will be slightly subluminal. We can simplify Eq.(5) by Taylor expanding $1-v_w=\varepsilon<<1$ to lowest order. Eq.(5) then reduces to

$$P_{analytic} \sim 2e^4B_y^2\gamma^2(\varepsilon + sin^2\alpha/2)^2/3m^2 \qquad (7).$$

This formula fully explains the physical origin of the result $P_{rad} << P_{syn}$. In relativistic PF acceleration, both $\varepsilon$ and $sin\alpha$ are $<<1$. Fig.9 shows an example for which the best-fit $\varepsilon = 0.03$. Eq.(7) shows that $P_{analytic}$ behaves differently depending on whether $\varepsilon >>$ or $<< sin^2\alpha/2$. In the former case $P_{analytic}$ depends only on the EM field profile speed $v_w$ and not on $\alpha$:

$$P_{analytic} \sim 2e^4B_y^2\gamma^2\varepsilon^2/3m^2 \qquad (8)$$

In the second case we regain Eq.(6) which depends only on $\alpha$ and not on $v_w$. Therefore when we model astrophysical data using these formulas, we obtain different physical information depending on the ratio $\varepsilon:\sin^2\alpha/2$, which depends on the PF initial condition, such as $T_o$, $\Omega_e/\omega_{pe}$ .etc. Eqs.(2), (6) & (8), which contain only 3 unknowns: $(B_y, \gamma, \alpha)$ or $(B_y, \gamma, \varepsilon)$, are much easier to use for modeling astrophysical data than Eq.(5) or Eq.(7), which contain 4 unknowns. In practice, $\varepsilon$ fluctuates rapidly in both time and space so its extraction from astrophysical data would be more difficult, whereas PIC simulation results suggest that $\sin\alpha$ has a narrower range for high-$\gamma$ particles (Fig.8). So the above analytic approximation is most useful for data modeling in the regime $\varepsilon \ll \sin^2\alpha/2$, which seems to be the case for GRB's (cf. Sec.7) and may also be the case for blazars.

We emphasize that the above analytic radiation formulas are derived from first principles independent of any PIC simulation results. Hence their validity is completely independent of any numerical simulation size scale. In fact we have carefully validated these analytic formulas using simulations spanning a dynamic range of $>10^5$ (i.e. PF pulse widths ranging from $10^2$ to $10^7$ gyroradii).

One may argue that the above result comes about only because we work in the lab frame, and that the classical synchrotron formula must apply if we transform to a (primed) Lorentz frame in which **E'** = 0 and **B'** is static. This is indeed true, and one can rederive the above formulas using appropriate Lorentz transformations of the classical synchrotron power formula (Rybicki and Lightman 1979). However, finding the Lorentz frame with **E'** = 0 is impractical, since E/B varies rapidly in both space and time due to modulation by self-generated currents and current instabilities (Liang and Nishimura 2004). There is no single Lorentz transformation that can lead to **E'** = 0 for any meaningful fraction of the EM pulse. So in practice it is more convenient to work in a global lab frame, measure **E, B** and **p** in this frame and use the above analytic formulas. Such lab-frame quantities are more relevant to astrophysical observations anyway.

## 6. IMPLICATIONS FOR ASTROPHYSICAL PF MODELS

Using PIC simulations and analytic methods, we have demonstrated above that electrons accelerated by Poynting flux which comoves with the local **ExB** drift velocity, radiate at a rate (in the lab-frame) much below the classical synchrotron power, and the critical frequency of their radiation spectrum is also much below the classical synchrotron critical frequency. These results have major implications for the interpretation of astrophysical data from GRBs and blazars, if their energy supply is coming from Poynting flux. One scenario would be that the classical synchrotron model applies only to the radiation zone, which is separate from the particle acceleration zone, and the synchrotron model (B, $\gamma$) values refer only to the radiation zone but not the acceleration zone, which likely has higher B and $\gamma$. An alternative scenario is that the observed radiation is intrinsic to the acceleration process. Then we have to use Eq.(5) or its various limits (Eq.(2)-(8)) to model the astrophysics data. This will lead to much higher (B, $\gamma$) values for the source than in conventional synchrotron models since $\varepsilon$ and $\alpha$ are <<1. Clearly the overall energetics and parameters of the two models will be very different and such differences may be testable. As an example of the application of the results of Secs. 4 & 5, we will consider a simplistic model of long GRBs in Sec.7. Another important consequence of the suppression of synchrotron radiation in PF scenarios is that inverse Comptonization of external soft photons (EC) may dominate even when the (lab-frame) magnetic energy density greatly exceeds the external soft photon energy density (Rybicki & Lightman 1979), and the conventional SSC + EC model of blazars (Dermer & Boettcher 2002) may need to be revamped. These scenarios will be reconsidered in future papers.

## 7. APPLICATION TO A PF MODEL OF LONG GRB's

Currently there is no universally accepted model of GRB energization and radiation. Two popular paradigms are hydrodynamic versus electromagnetic outflows from a central engine (e.g. a newly formed black hole), dissipating at a distance of $10^{14-15}$ cm via nonthermal electrons and gamma-rays (Meszaros 2002, Piran 2000). If GRBs are indeed the manifestations of intense PF outflow, a dissipation mechanism such as TPA is attractive due to its high energy conversion efficiency and power-law spectra (cf. Sec.2). Liang and Nishimura (2004) also pointed out several tantalizing similarities between the observable properties of TPA in e+e- plasmas and GRB phenomenology. To demonstrate the utility of the results of Secs.4&5, here we apply the analytic radiation formulas to a simplistic "toy" model of classical long GRBs, assuming that the PF contains only e+e- pairs with no ions (e-ion models will be considered in subsequent papers). We will derive the value of the spectral break energy $E_{pk}$ from these formulas. The underlying astrophysical framework is that some central engine activity lasting 10's of seconds launches an intense EM pulse of width $\sim 10^{12}$ cm and energy $\sim 10^{51}$ ergs, loaded with only low-density e+e- plasma so that $\Omega_e/\omega_{pe} > 1$. This intense EM pulse initially propagates through the collapsar envelope as a non-dissipative subluminal MHD pulse as long as the ambient density is high enough so that the formal Alfven speed $v_A = B/(4\pi\rho_p)^{1/2} < c$ ($\rho_p$ = ambient proton mass density). But the pulse will eventually reach a point where the envelope ion density is so low that $v_A > c$, and the MHD pulse turns into a freely-expanding kinetic EM pulse. This triggers the TPA dissipation and rapid conversion of EM energy into e+e- kinetic energy. We have performed simulations of relativistic magnetosonic pulses propagating down steep density gradients. The preliminary results seem to support the above picture.

For long GRBs it is useful to scale the burst parameters with the following benchmark values (Fishman & Meegan 1998, Preece et al 2000): total energy $E_{51} = E_{tot}/10^{51}$erg, burst duration $T_{30} = T_{90}/30$sec, prompt-$\gamma$ emission distance $R_{14} = R/10^{14}$cm. We assume that the EM pulse is a quasi-spherical shell with thickness $\Delta R = cT_{90} = 10^{12}$cm$T_{30}$ (in this section we write out c explicitly) and solid angle $\Omega_{4\pi} = \Omega/4\pi$. To simplify the model we assume that the shell is uniform with mean field B and mean lepton (e- + e+) density n. *All physical quantities are measured in the "lab-frame", which we assume to be the rest frame of the GRB central engine or host galaxy.* In reality, the field, density and momentum profiles are highly structured due to current instabilities (Liang and Nishimura 2004), and the following parameters refer primarily to those leptons at the peak of the momentum distribution function. All our simulations with pair plasmas suggest that at late times, particle energy $E_{particle} \geq 0.6E_{tot}$, EM energy $(= 2E_B) \leq 0.4E_{tot}$ (Fig.10, see also Liang et al 2003). Let N = total number of leptons (e+ + e-) in the pulse and $\Gamma$ = average Lorentz factor of the lepton distribution = $<\gamma f(\gamma)>$. ($\Gamma \sim$ the group veocity Lorentz factor of the EM pulse $\Gamma_w = (1 - v_w^2)^{-1/2}$, Liang and Nishimura 2004). We thus have dimensionally in cgs units:

$$N\Gamma mc^2 \sim 6 \times 10^{50} E_{51} \tag{9}$$

$$B^2 \Delta R R^2 \Omega \sim 16\pi \times 10^{50} E_{51} \tag{10}$$

Eq.(10) gives:

$$B \sim 2 \times 10^5 \text{ G } (R_{14}^{-1} \Omega_{4\pi}^{-1/2} E_{51}^{1/2} T_{30}^{-1/2}) \tag{11}$$

Next we estimate $\Gamma$ for the radiation epoch by invoking the condition: radiative cooling rate = particle acceleration rate. Liang and Nishimura (2004) derived from the Lorentz force equation the particle acceleration rate $d\Gamma/dt = f\Omega_c/\Gamma$, where f is a fudge parameter of O(1) that depends on the initial conditions. We emphasize that this formula depends solely on the comoving nature of the EM pulse and is independent of other global assumption of the TPA. Using Eq.(7) and assuming $\varepsilon \ll \sin\alpha$ (and check for consistency below), we obtain:

$$f\, ecB/\Gamma = e^4 B_y^2 \Gamma^2 \sin^4\alpha/6m^2c^3 \qquad (12).$$

The peak value of the late-time $\sin\alpha$ distribution for high-$\gamma$ electrons (Fig.8) lies in the range 0.01 to 0.2 in the simulations performed so far. Hence we scale $\sin\alpha$ with 0.1 below: $\alpha_{.1} = \sin\alpha/0.1$. Solving Eq.(12) for $\Gamma$ we obtain:

$$\Gamma \sim 1.2 \times 10^5\ (f^{1/3}\, R_{14}^{1/3}\, \Omega_{4\pi}^{1/6}\, E_{51}^{-1/6}\, T_{30}^{1/6}\, \alpha_{.1}^{-4/3}) \qquad (13)$$

Hence $\varepsilon \sim 1/\Gamma \ll \sin\alpha$ and our assumption is justified. Note that this $\Gamma$ is measured in the lab-frame so it is actually smaller than the composite Lorentz factor of internal shock models (with a bulk Lorentz factor of $10^2$ times internal Lorentz factors of $10^3$-$10^4$). Using this in Eq.(9) we find:

$$N \sim 6 \times 10^{51}\ (f^{-1/3}\, R_{14}^{-1/3}\, \Omega_{4\pi}^{-1/6}\, E_{51}^{7/6}\, T_{30}^{1/6}\, \alpha_{.1}^{4/3}) \qquad (14)$$

From Eqs.(3), (11) and (13) we obtain the value of the spectral break energy, taken as the critical frequency corresponding to $\Gamma$:

$$E_{pk} = h\omega_{cr}/2\pi \sim 490\ \text{keV} (f^{2/3}\, R_{14}^{-1/3}\, \Omega_{4\pi}^{-1/6}\, E_{51}^{1/6}\, T_{30}^{-1/6}\, \alpha_{.1}^{-2/3}) \qquad (15)$$

This value agrees with the observed spectral breaks of typical long GRBs: $E_{pk} \sim 250$ keV $(1+z) \sim 500$ keV for $z \sim 1$ (Preece et al 2000) in the host-Galaxy frame. Note that $E_{pk}$ in Eq.(15) depends only weakly on the various uncertainty factors. Eq.(14) gives the mean lepton density:

$$n = N/(\Omega \Delta R R^2) \sim 5 \times 10^{10} (f^{-1/3}\, R_{14}^{-7/3}\, \Omega_{4\pi}^{-1/6}\, E_{51}^{7/6}\, T_{30}^{-7/6}\, \alpha_{.1}^{4/3}) \qquad (16)$$

and the frequency ratio:

$$\Omega_c/\omega_{pe} \sim 250\ (f^{1/6}\, R_{14}^{1/6}\, \Omega_{4\pi}^{-5/12}\, E_{51}^{-1/12}\, T_{30}^{1/12}\, \alpha_{.1}^{-2/3}) \gg 1 \qquad (17)$$

which justifies our EM-domination assumption. At this density the pairs are completely collisionless (Coulomb mean free path $> 10^{20}$cm). We note that the local acceleration (=cooling) time of an individual electron with the above B, Γ and sinα values is very short: $t_{cool} = t_{cool} \sim 10^{-2}$ sec. However, this should *not* conflict with the 30 second observed GRB duration. The radiation duration of the shell is determined by the conversion time of overall EM energy into particle energy, which is proportional to the light crossing time of the shell thickness ΔR/c, since the EM pulse takes at least that long to emerge and energize the embedded plasma (Fig.10). Moreover, radiation emitted by the front and back of the plasma arrive at the detector with a time delay of ΔR/c. These two effects combine to make the GRB duration measured by the detector ~ ΔR/c = 30 sec, irrespective of the short acceleration/cooling time of individual leptons. An analogy with internal shock models is in order here. Even though the internal shock thickness and particle acceleration length are << $10^{12}$ cm, the GRB radiation duration is governed by the shock crossing time of the colliding shells, which have thicknesses of ~ $10^{12}$ cm. We note that $10^{12}$cm corresponds to ~$10^{14}$ gyroradii. Even the acceleration/cooling distance of ~$10^8$cm equals $10^{10}$ gyroradii. Both scales are much larger than the largest PIC simulation we have performed (~$10^7$ gyroradii). Hence it is tempting to question the applicability of our results to the GRB regimes. But we emphasize that the only physics used to derive the radiation rate Eq.(7) and the acceleration rate of Liang and Nishimura (2004) is the comoving assumption of the local EM wave with the local **ExB** drift speed. This assumption is completely independent of the global geometry, structure and size scale of the fields and plasmas, which affect only the duratiion and longevity of the acceleration process. In addition, we have validated the analytic radiation and acceleration rates with simulations spanning over 4 decades in physical size ($10^3$ to $10^7$ gyroradii). This gives us added confidence in their general validity.

However one puzzle remains. What makes the EM pulse dissipation to occur at R~$10^{14}$ cm from the central engine, two orders of magnitude larger than the EM pulse width and six orders of magnitudes larger than the lepton acceleration/cooling length? We speculate that it may be the environment which determines this dissipation distance. Here we venture a somewhat speculative but plausible scenario that gives rise to such a large dissipation distance. In reality, the TPA action takes place not at a sharp boundary but in an external density gradient whose scale height is much larger than the acceleration length. In such cases we believe that the Liang-Nishimura (2004) acceleration rate and the radiation cooling rate of Sec.5 are applicable only when the ambient ion mass density drops below the internal pair mass density. Otherwise the EM expansion and particle acceleration would be inhibited by the external ion inertia (cf. discussion at the beginning of Sec.7). In the collapsar model the GRB progenitor is likely surrounded by a Wolf-Rayet wind whose mass density $\sim$ A $5 \times 10^{11}$ r$^{-2}$ g.cm$^{-1}$ (Chevalier and Li 2000) where the parameter A depends on the mass loss rate. Hence the PF "breakout" or dissipation distance, using the pair density of Eq.(16), becomes $r_{breakout} \sim A^{1/2} 10^{14}$ cm. In other words, the TPA action is inhibited until the PF reaches an ambient ion mass density of $\leq 5 \times 10^{-17}$ g.cm$^{-3}$, and this can only occur at a distance $\geq 10^{14}$ cm in a Wolf-Rayet wind. Fig.11 illustrates the relevant length scales of this scenario.

## 8. DISCUSSION AND SUMMARY

Using PIC simulations and analytic theory, we have shown in this paper that when electrons are accelerated by a comoving Poynting flux with $\Omega_c/\omega_{pe} > 1$, the in-situ radiation power output and critical frequency are much lower than those given by the classical synchrotron formulas. This is because the most energetic electrons have their momentum closely aligned with the local Poynting vector or **ExB** drift direction. We apply our analytic formulas for the radiation power output and critical frequency to a simple PF model of classical long GRBs, and find that the predicted spectral break energy lies in the range of observed data.

Besides the LPA and TPA mechanisms which involve ponderomotive acceleration of overdense plasmas, there are many other Poynting flux scenarios that may result in nonthermal particle acceleration. For example, electron acceleration by comoving longitudinal wakefields generated by PF in an underdense plasma (similar to laser accelerators in the laboratory, Tajima and Dawson 1979) may occur in special astrophysical situations. We have also not considered Poynting flux propagating along flow-aligned guide fields such as Alfven and whistler waves. Preliminary PIC simulation results suggest that linear Alfven waves ($\Delta B << B_o$ where $B_o$ is the longitudinal guide field) cannot accelerate nonthermal particles efficiently via the ponderomotive force, since the net **E x B** drift direction is misaligned from the (strong) guide field. On the other hand, if the Alfven wave is highly nonlinear ($\Delta B > B_o$), it behaves like transverse EM waves. Then the TPA results may apply to first order. Nonlinear Alfven waves also couple to longitudinal modes via parametric decay, and the Langmuir waves can then accelerate the electrons. In general, waves can transfer energy to electrons via a large variety of resonant interactions (Boyd and Sanderson 1969). But such resonant interactions act on only a small population of the electrons infrequently, whereas the ponderomotive force can accelerate the bulk of the plasma. PF acceleration in e-ion plasmas is more complex than in e+e- plasmas due to charge separation (Nishimura et al 2003). Their radiation will be treated in a separate paper.

This work was partially supported by NSF AST0406882 and NASA NNG06GH06G.

1. Rice University, Houston TX 77005-1892.

Figure Captions

Fig.1 Sample PIC simulation outputs illustrating the two different mechanisms of kinetic Poynting flux acceleration of an e+e- plasma slab: (a) in LPA, an intense EM pulse incident on an overdense plasma interface induces a ponderomotive force (**JxB** is along Poynting vector **k**) that snowplows relativistically all upstream electrons which must share the Poynting flux momentum, thus limiting their Lorentz factor; (b) in TPA, an intense EM pulse escaping from an overdense plasma induces a ponderomotive force which pulls out the surface electrons relativistically.  Only the fastest electrons can keep up with the EM pulse, so that the plasma loading of the EM pulse decreases with time.  TPA leads to the sustained comoving acceleration of a decreasing number of fast electrons, with no limit to their Lorentz factor.  In all figures of this paper, x is expressed in units of $3c/\omega_{pe}$.

Fig.2 (a) Electron energy spectrum accelerated by LPA resembles a superthermal quasi-Maxwellian distribution; (b) electron energy spectra accelerated by TPA for different initial PF thicknesses ($10^3$ and $10^4$ $c/\omega_{pe}$) both show a robust power-law of index $\sim$ -3 to -4.  The low-energy spectral breaks correspond roughly to the Lorentz factor of the EM pulse group velocity .

Fig.3  Calibration of the numerical radiation power $P_{rad}$ computed from the PIC simulation (Eq.(1)) against the analytic synchrotron formula $P_{syn}$ for a 5 MeV thermal plasma in a static uniform B field shows excellent agreement.  The scatter at low energies is due to small errors from interpolating the field values to the particle position.  In all figures of this paper, $P_{rad}$, $P_{syn}$ and $P_{analytic}$ are expressed in units of $2e^2\Omega_e^2/2700$.

Fig.4  Upper panel: Snapshots of $P_{rad}$ distribution (dots) and $B_y$ profile (solid, in units of $B_o/15$) vs. x for an e+e- plasma slab initially located at x=180 with n=16$n_{cr}$, thickness = 12$c/\omega_{pe}$ and snowplowed by a vacuum EM pulse with $\Omega_e/\omega_{pe}$=10 from left to right.  $P_{rad}$ of the accelerated

electrons reaches a maximum at ~2 light crossing times after the EM pulse hits the plasma surface, followed by rapid monotonic decay. This behavior is caused by the decrease in angle $\alpha$ competing with the increase in electron energy. Lower panel: Snapshots of the critical frequency (Eq.3) distribution show that the evolution of the spectral hardness of radiation follows that of $P_{rad}$. $\omega_{cr}$ is expressed in units of $10\Omega_e$.

Fig.5 Upper panel: Snapshots of $P_{rad}$ distribution (dots) and $B_y$ profile (solid, in units of $B_o/150$) vs. x for a e+e- plasma slab accelerated by TPA with initial plasma temperature $kT_o=0.005m$, thickness $L_o=12c/\omega_{pe}$, $\Omega_e/\omega_{pe}=10$ and initially located at x=180. $P_{rad}$ of the accelerated electrons reaches a maximum at ~5 light crossing times after the emergence of the EM pulse, followed by monotonic decay which is slower than in the LPA case. Lower panel: Snapshots of the critical frequency (Eq.3) distribution shows that the evolution of the spectral hardness of radiation follows that of $P_{rad}$. $\omega_{cr}$ is expressed in units of $10\Omega_e$. Note that $\omega_{cr}$ of Fig.5 is a factor of 10 larger than that of Fig.4. The radiation at t=0 is thermal cyclotron radiation due to the finite initial temperature. But it has very low $\omega_{cr}$

Fig.6 Diagram showing the angle $\alpha$ between the Poynting vector **k** and **p$_+$,** the momentum component orthogonal to **B**.

Fig.7 Scatter plot of $P_{rad}$ compared with $P_{analytic}$ of Eq.(2) for the runs of (a) Fig.4 and (b) Fig.5. At these times most of the high–$\gamma$ particles are comoving with the EM pulse.

Fig.8 Scatter plot of the distribution of Lorentz factor $\gamma$ vs. $\sin\alpha=p_z/\gamma$ for a $\Omega_e/\omega_{pe}=10$ e+e- slab accelerated by TPA shows that the highest–$\gamma$ particles have their $\sin\alpha$ distribution peaking in the range ~ 0.01- 0.2.

Fig.9 Scatter plot of $P_{rad}$ compared with $P_{analytic}$ of Eq.(7) for an $\Omega_e/\omega_{pe}=10$ e+e- slab accelerated by TPA, At this time the best correlation for high-$\gamma$ electrons is obtained when $\epsilon = 0.03$.

Fig.10  Decay curves of EM energy for an $\Omega_e/\omega_{pe}=10$, $kT_o=5MeV$ e+e- slab TPA's with different initial thicknesses: (A) $L_o=10800c/\omega_{pe}$; (B) $L_o=90c/\omega_{pe}$; (C) $L_o=12c/\omega_{pe}$.  This confirms that the conversion time of EM energy into particle energy is directly proportional to the light crossing time $L_o/c$.

Fig.11  Diagram illustrating the different size scales in the "breakout" of a PF from a Wolf-Rayet wind model of long GRBs.  The wavy arrow denotes the (lab-frame) PF thickness ($\Delta R=10^{12}cm$) along the observer line of sight.  The PF breakout distance ($\sim 10^{14}\,cm$) is determined by the radius at which the wind mass density drops below the PF pair mass density ($\sim 5x10^{-17}\,g.cm^{-3}$).  Despite the short acceleration/cooling length ($\sim 3x10^{8}cm$) of individual leptons accelerated by the PF, the detector-measured GRB duration at infinity is $\sim\Delta R/c=30$ sec due to the transit time of the PF crossing $r_{breakout}$ and the light path difference between the front and back of the PF (upper-right space-time diagram).

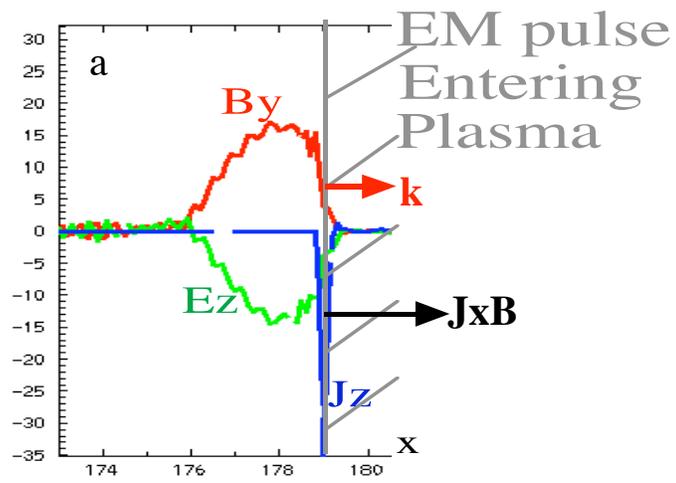

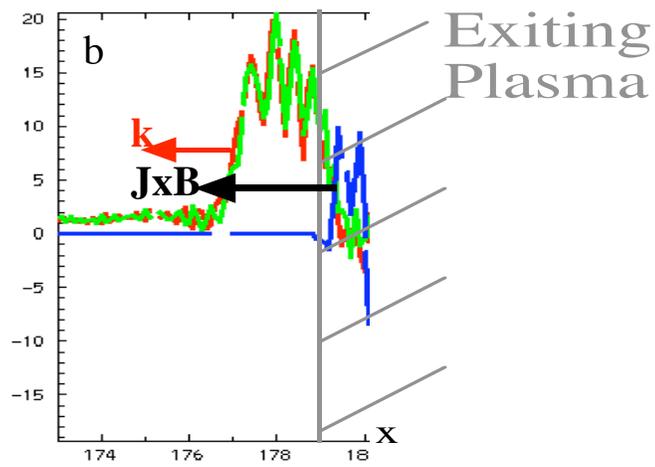

Fig.1

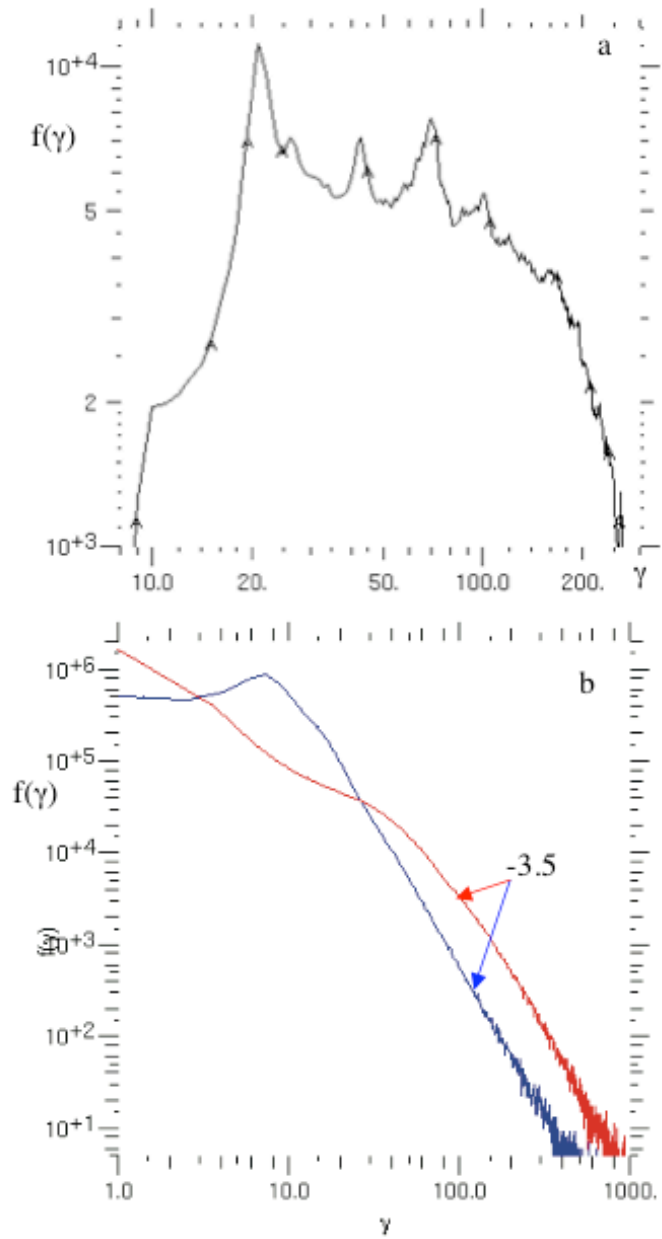

Fig.2

Fig.2

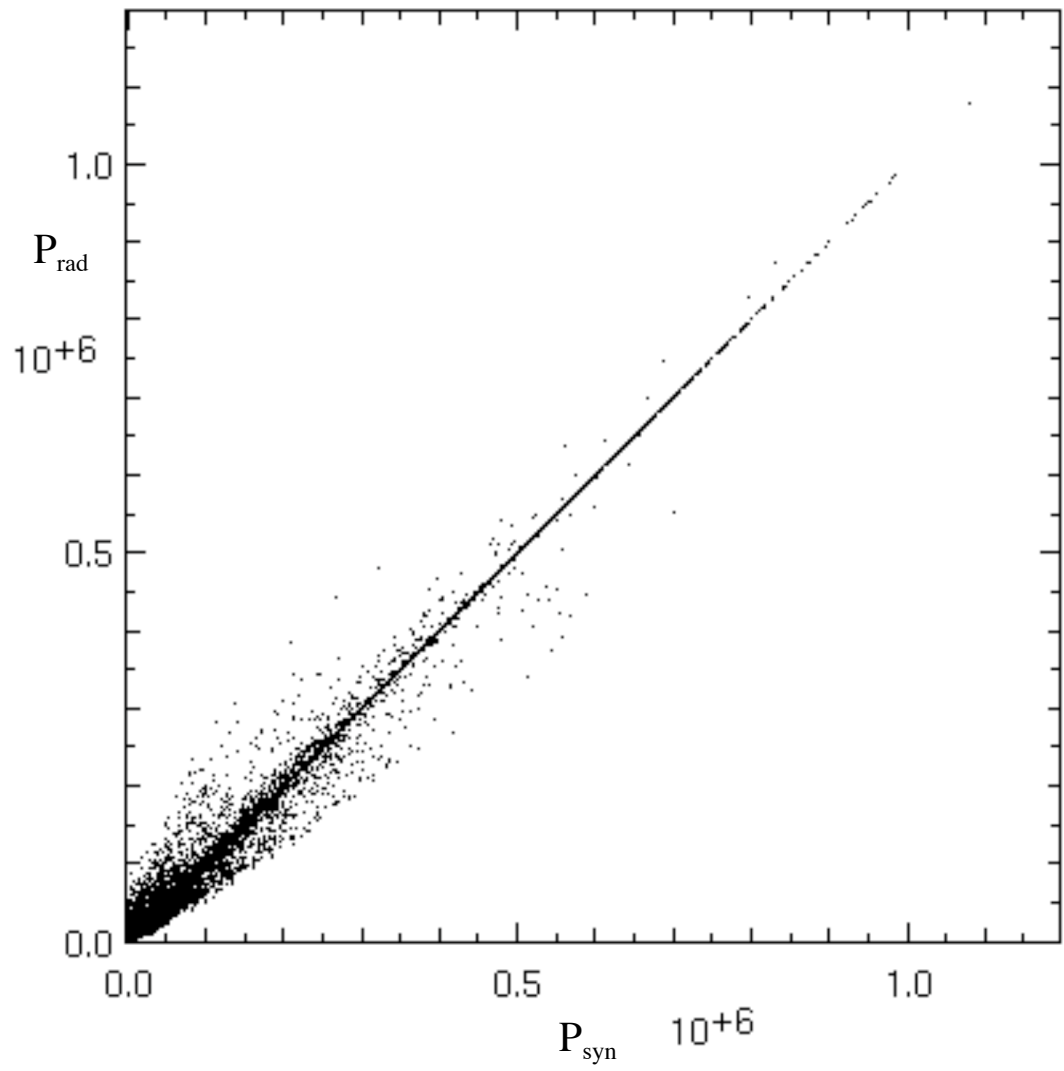

Fig.3

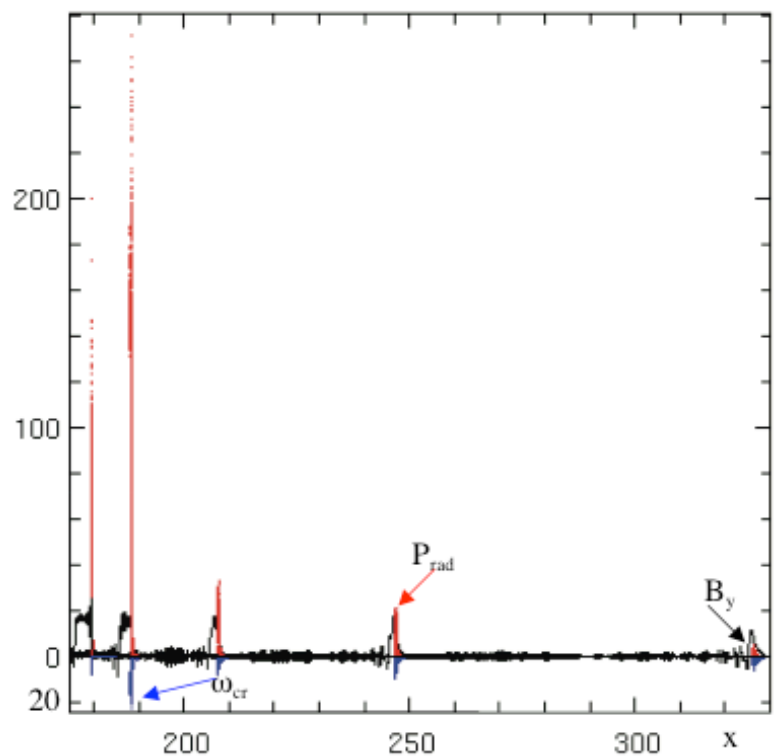

Fig.4

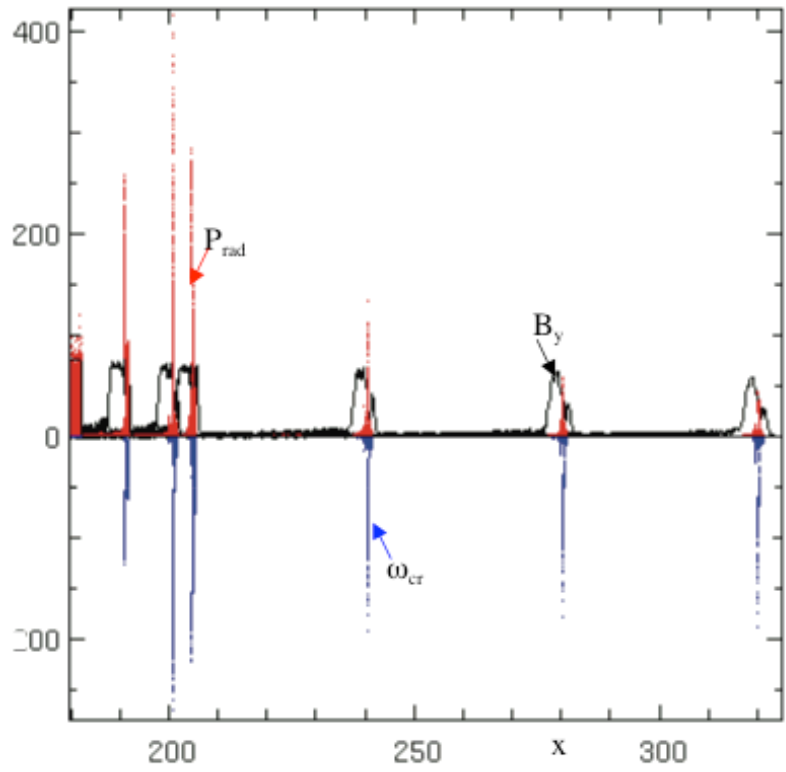

Fig.5

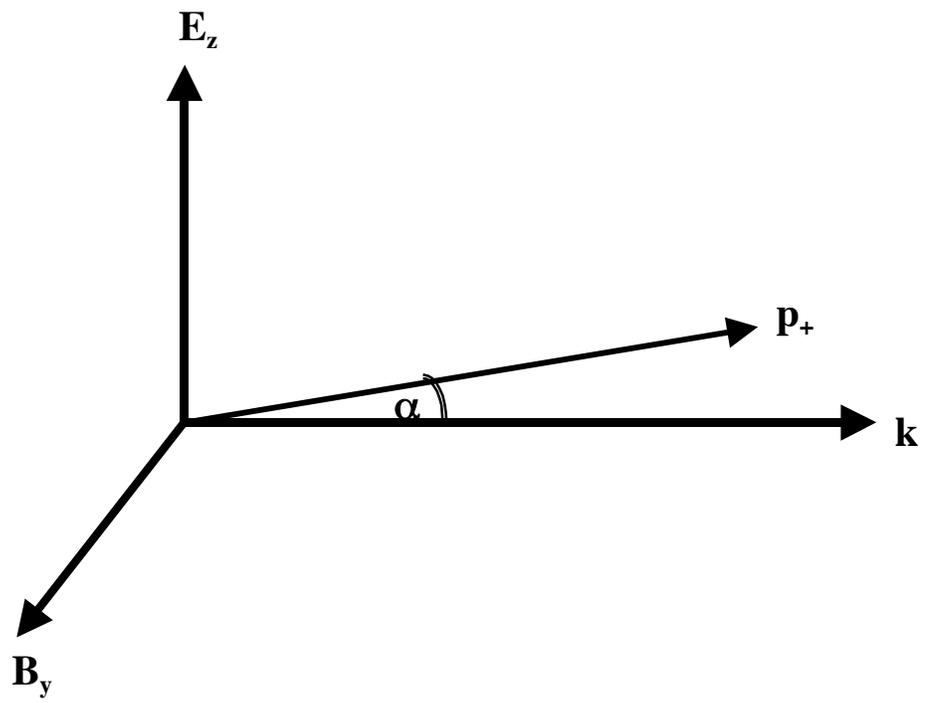

Fig.6

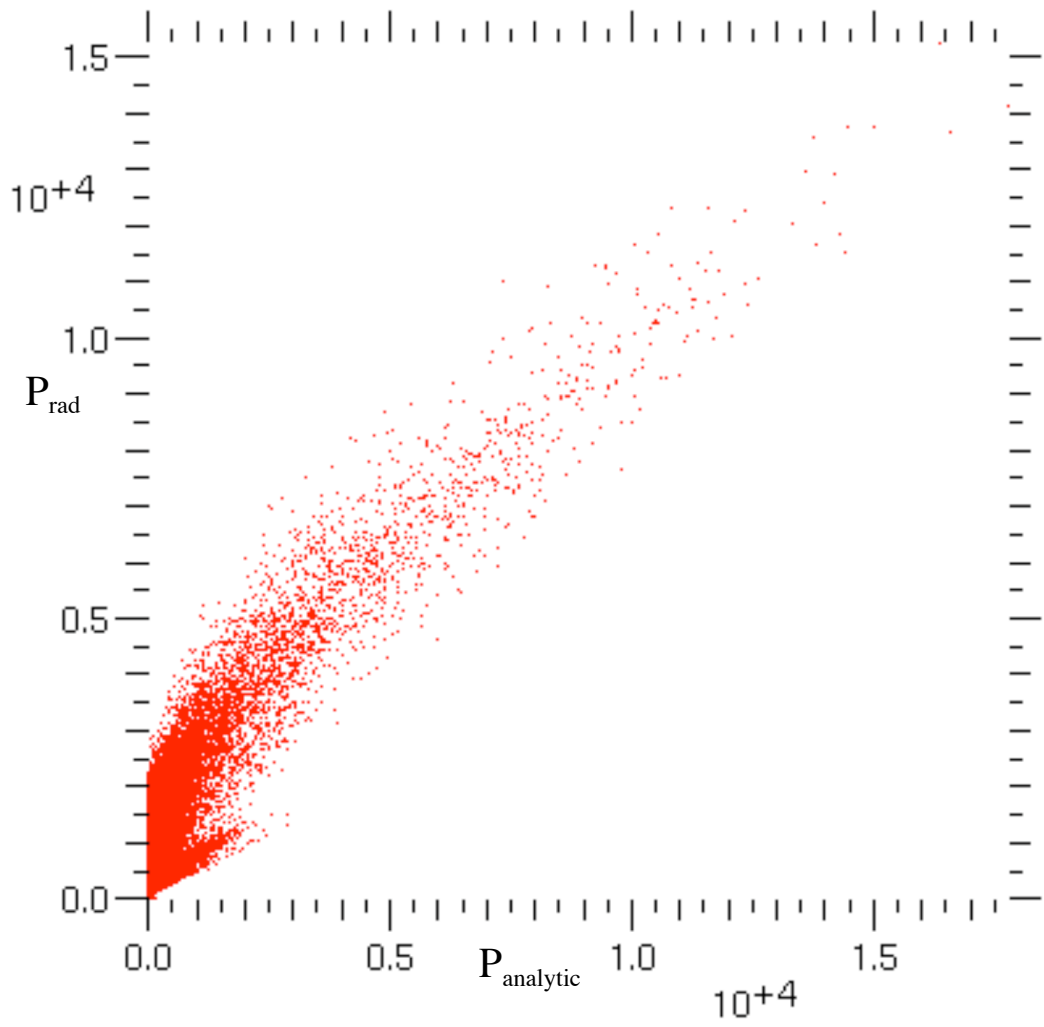

Fig.7

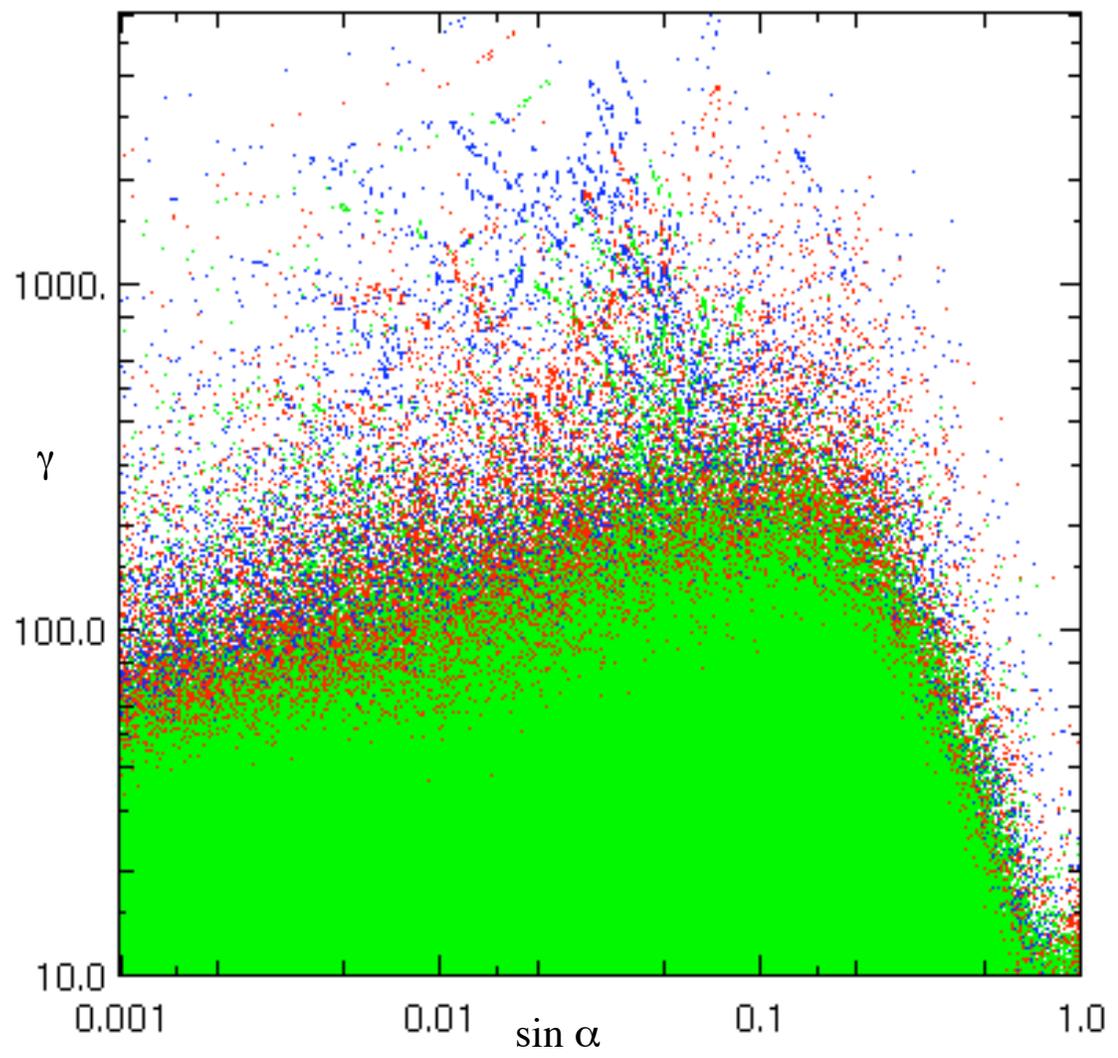

Fig.8

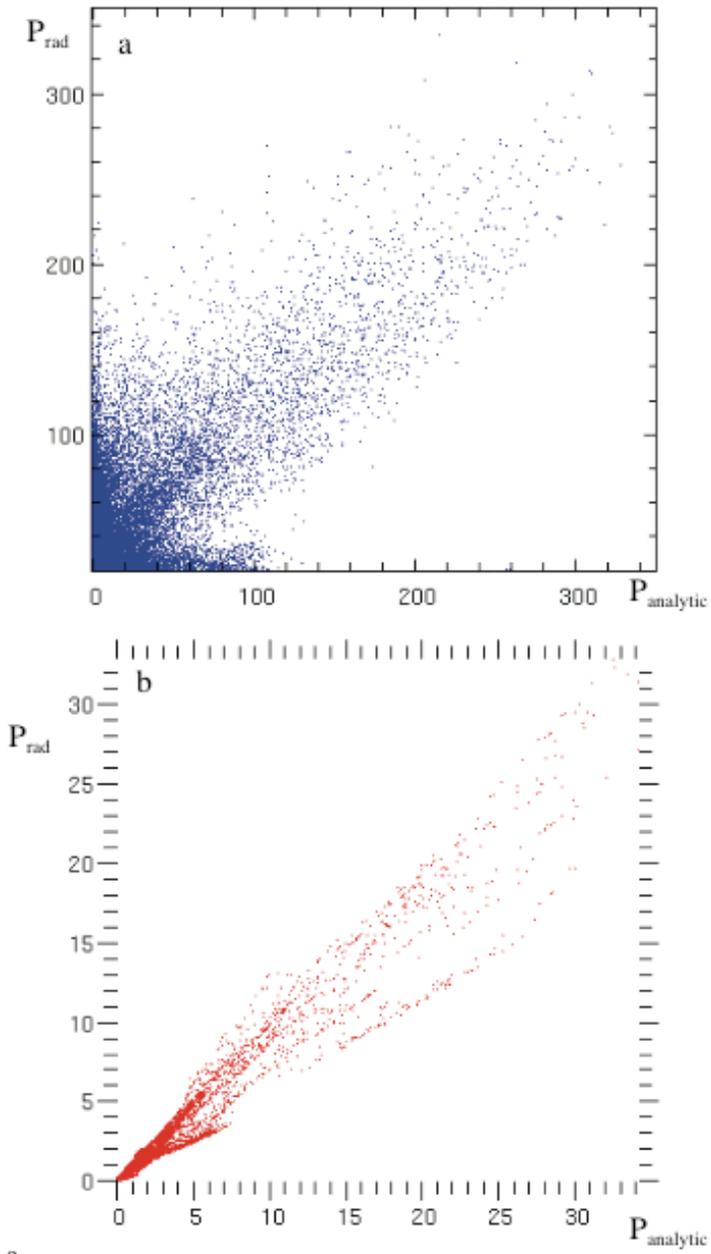

Fig.9

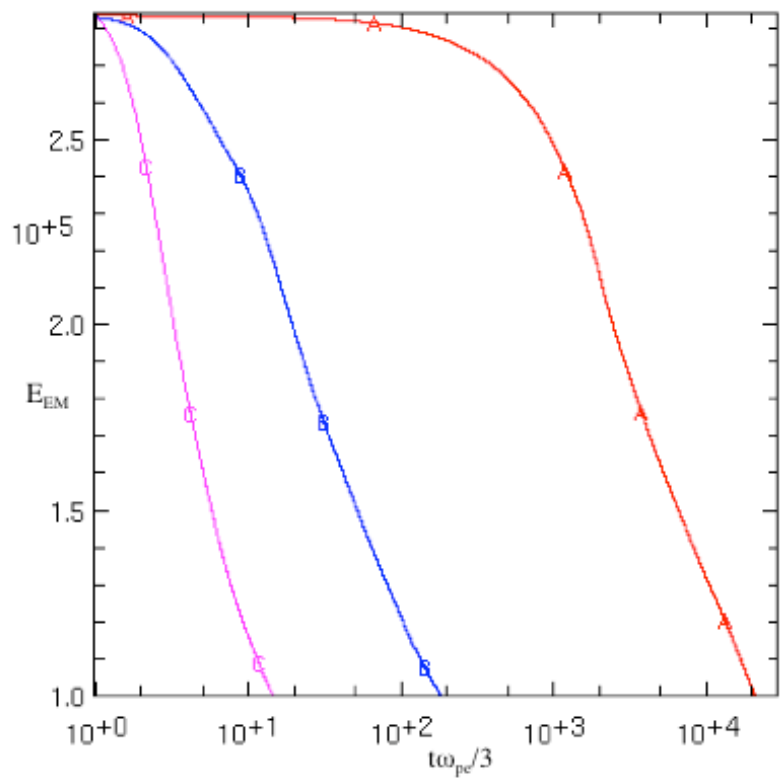

Fig.10

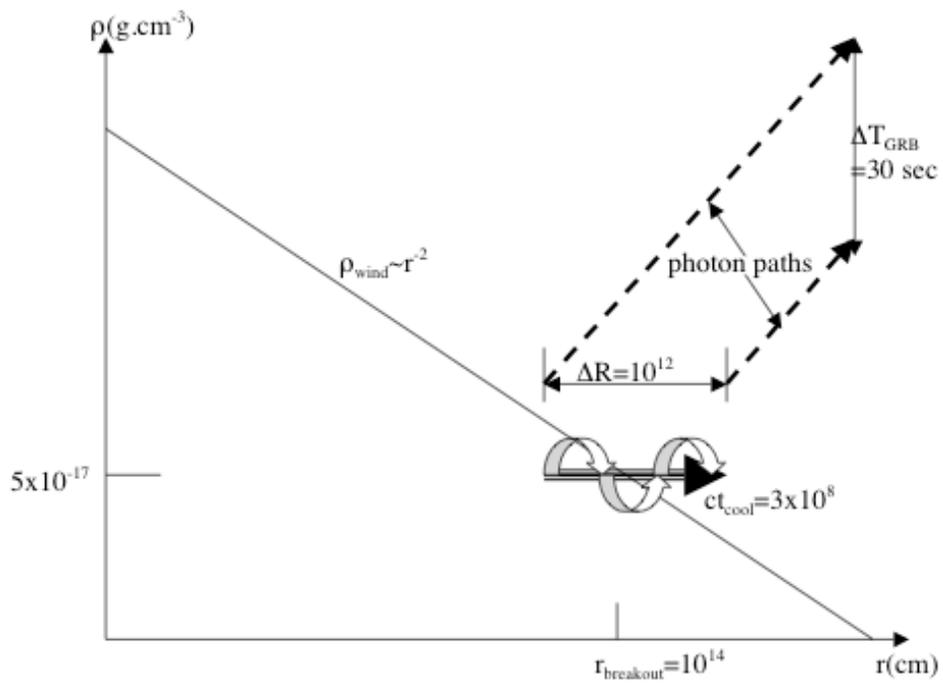

Fig.11